\documentclass{article}
\usepackage{amssymb}

\usepackage{graphicx}
\usepackage{amsmath}

\input{tcilatex}

\begin{document}

\title{Arbitrary amplitude inertial Alfv\'{e}n waves in homogeneous magnetized
electron-positron-ion plasmas}
\author{S. Mahmood and H. Saleem \\
Department of Physics, COMSATS Institute of Information Technology,\\
H-8 Islamabad Pakistan\\
and\\
PINSTECH (Physics Research Division) P.O. Nilore\\
Islamabad Pakistan}
\maketitle

\begin{abstract}
Nonlinear set of equations for inertial or slow shear Alfv\'{e}n wave (SSAW)
in ideal electron-positron-ion (e-p-i) plasmas are presented. The analytical
solution for arbitrary amplitude SSAW in such multi-component plasmas is
obtained using Sagdeev potential approach. The numerical solutions for
several different cases have also been presented for illustrative purpose.
It is found that the electron density dips of SSAW are formed in the super
Alfv\'{e}nic region. The amplitude and the width of the nonlinear shear Alfv%
\'{e}n wave reduces with the increase in the concentration of positrons in
electron-ion (e-i) plasmas. The width of the soliton also depends upon the
direction of propagation of the perturbation in both e-i and e-p-i
plasmas.\newpage 
\end{abstract}

\begin{center}
{\Large I. INTRODUCTION\bigskip }
\end{center}

The solitary kinetic Alfv\'{e}n wave (KAW) was studied in electron-ion (e-i)
plasmas by Hasegawa and Mima \cite{r1} long ago. These waves can propagate
in moderate-$\beta $ plasmas for $\frac{m_{e}}{m_{i}}<\beta (=\frac{%
4\barwedge nT}{B_{0}^{2}})<1$. The phase velocity of the KAW is less than
the thermal speed of the electrons and hot electrons are assumed to be
inertialess and follow the Boltzmann distribution. Later on the nonlinear
coupling of KAW and ion acoustic wave (IAW) in e-i plasma was studied by Yu
and Shukla \cite{r2}

In low-$\beta $ ($<\frac{m_{e}}{m_{i}}$) plasma, the electron inertia cannot
be ignored and the wave propagation is slowed down. The phase velocity of
Inertial or slow shear Alfv\'{e}n wave (SSAW) is larger than the thermal
speed of the electrons. The nonlinear (SSAW) was studied in e-i plasma with $%
\beta <\frac{m_{e}}{m_{i}}$ about two decades ago \cite{r3} using Sagdeev
potential approach. The conditions for the existence of localized solitary
solutions were discussed. The linear dispersion relation of SSAW in e-i
plasmas is $\omega ^{2}=\frac{v_{Ai}^{2}k_{z}^{2}}{(1+\lambda
_{e}^{2}k_{\bot }^{2})}$ where the external magnetic field is $\mathbf{B}%
_{0}=B_{0}\hat{z}$, Alfv\'{e}n speed is $v_{Ai}=(\frac{B_{0}^{2}}{4\barwedge
n_{i0}m_{i}})^{\frac{1}{2}}$ (here $n_{i0\text{ }}$and $m_{i}$ are the ion
unperturbed density and mass, respectively) and $\lambda _{e}=\frac{c}{%
\omega _{pe}}$ ( here $c$ is the speed of light and $\omega _{pe}=\left( 
\frac{4\barwedge n_{0}e^{2}}{m_{e}}\right) ^{\frac{1}{2}}$ is the electron
plasma frequency) is the electron collisionless skin depth in e-i plasmas.
Furthermore $k_{z}$ and $k_{\perp }$ are the parallel and perpendicular wave
vectors, respectively, with respect to $\mathbf{B}_{0}$.

However, the KAW has faster phase velocity than SSAW and its linear
dispersion relation in e-i plasmas is $\omega
^{2}=v_{Ai}^{2}k_{z}^{2}(1+\rho _{s}^{2}k_{\perp }^{2})$ , where $\rho _{s}=%
\frac{c_{s}}{\Omega _{i}}$ is the ion Larmor radius at electron temperature
and $c_{s}=\left( \frac{T_{e}}{m_{i}}\right) ^{\frac{1}{2}}$ is the ion
sound speed ). The ion inertia plays an important role in shear Alfv\'{e}n
wave dynamics due to ion polarization drift under low frequency assumption
i.e., $|\partial _{t}|<<\Omega _{i}$ (where $\Omega _{i}=\frac{eB_{0}}{m_{i}c%
}$ is the ion-gyrofrequency).

During the last decade, there has been a great deal of interest in
electron-positron-ion \ (e-p-i) plasmas and a number of authors \cite
{r4,r5,r6,r7,r8,r9,r10} have studied different linear and nonlinear wave
propagation phenomenon in such sytems. The e-p-i plasmas are supposed to
exist in active galactic nuclei (AGN), pulsar magnetospheres and early
universe etc.\cite{r5,r6,r7,r8,r9,r10}. The positrons are introduced in e-i
plasmas for the purpose of diagnostics and to model the pulsar magnetosphere
in laboratory experiments \cite{r11,r12,r13} In this paper, we show that the
presence of positrons in e-i plasmas can change the nonlinear dynamics of
low frequency slow shear Alfv\'{e}n wave significantly. Or we may say the
other way round, that presence of ions in e-p plasmas can change both the
spatial and temporal scales. Therefore it seems important to study the
linear as well as nonlinear phenomenon in e-p-i plasmas to explain some
aspects of laboratory or astrophysical e-p-i plasmas.

Recently kinetic Alfv\'{e}n wave (KAW) has been studied in
electron-positron-ion (e-p-i) plasmas by Saleem and Mahmood \cite{r10}. In
this case, the electrons and positrons have been assumed to be inertialess
and they follow the Boltzmann distribution. The ions are considered to be
inertial and cold. The phase velocity of the wave is less than the thermal
speed of both the electrons and positrons. The polarization drifts of
electrons and positrons can be ignored in e-p-i plasmas and the ion
polarization drift is necessary to maintain quasineutrality conditions.

The paper has been presented in this manner. In Sec. II nonlinear equations
for Inertial or SSAW in e-p-i have been defined. The localized stationary
solution in the form of energy integral equation has been obtained in Sec.
III. Some of the possible numerical solution are presented in Sec. IV. In
Sec. V discussion on the obtained results is presented.\bigskip

\begin{center}
II. {\Large SET\ OF\ EQUATIONS}
\end{center}

Let us consider a cold electron-positron-ion (e-p-i) plasma in the presence
of external magnetic field $\mathbf{B}_{0}=B_{0}\hat{z}$ along z-axis. The
governing equations for the nonlinear Alfv\'{e}n wave dynamics in x-z plane
in a low $\beta $ (i.e., $\beta <<\frac{m_{e}}{m_{i}}$) e-p-i plasma with $%
v_{tj}<\frac{\omega }{k_{z}}$ (where $v_{tj}^{2}=\frac{T_{j}}{m_{j}}$ and $%
j=e,p$) are as follows:

The continuity equations for electrons and positrons can be written as, 
\begin{equation}
\partial _{t}n_{j}+\partial _{z}(n_{j}v_{jz})=0  \label{e1}
\end{equation}
whereas the equations of motion for electrons and positrons along \^{z}-axis
are, 
\begin{equation}
\partial _{t}v_{jz}+v_{jz}\partial _{z}v_{jz}=\frac{q_{j}}{m_{j}}E_{z}
\label{e2}
\end{equation}
where $j=e,p$ and $q_{j}=-e,+e$ for electrons and positrons, respectively.

The ion equation of motion in the limit $|\partial _{t}|<<\Omega _{i}$ can
be expressed as, 
\begin{equation}
v_{ix}=\frac{c}{B_{0}\Omega _{i}}\partial _{t}E_{x}  \label{e5}
\end{equation}
We are ignoring the ion parallel motion along the magnetic field, so the ion
continuity equation gives, 
\begin{equation}
\partial _{t}n_{i}+\frac{c}{B_{0}\Omega _{i}}\partial _{x}\left(
n_{i}\partial _{t}E_{x}\right) =0  \label{e6}
\end{equation}
The Faradays law $\mathbf{\nabla \times E=-}\frac{1}{c}\partial _{t}\mathbf{B%
}$ can be written as, 
\begin{equation}
\partial _{z}E_{x}-\partial _{x}E_{z}=-\frac{1}{c}\partial _{t}B_{y}
\label{e7}
\end{equation}
and Ampere's law yields, 
\begin{equation}
\partial _{x}B_{y}=\frac{4\pi e}{c}\left( n_{p}v_{pz}-n_{e}v_{ez}\right)
\label{e8}
\end{equation}
where the displacement current has been ignored.

The quasi-neutrality condition implies, 
\begin{equation}
n_{i}\simeq n_{e}-n_{p}  \label{e9}
\end{equation}
where $n_{\alpha }$ (where $\alpha =e,p,i$) are the densities, $v_{\alpha z}$
and $v_{\alpha x}$ are the parallel and perpendicular velocities of $\alpha $%
-species with respect to the external magnetic field, respectively.

Using two potential approach, one can express the parallel and perpendicular
electric fields as $E_{x}=-\partial _{x}\phi $ and $E_{z}=-\partial _{z}\psi 
$, where \ `$\phi $' and `$\psi $' are electrostatic and electromagnetic
potentials, respectively.

The linear dispersion relation of inertial Alfv\'{e}n wave in low-$\beta $
e-p-i plasmas turns out to be, 
\begin{equation}
\omega ^{2}=\frac{\left( 1+p\right) v_{Ai}^{2}k_{z}^{2}}{\left[ \left(
1+p\right) +\lambda _{e}^{2}k_{x}^{2}\right] }  \label{e10}
\end{equation}
where $p=\frac{n_{p0}}{n_{e0}}$ (here $n_{p0}$ and $n_{e0}$ are the
unperturbed densities of positrons and electrons, respectively).

The limiting case of two component e-i plasmas can be obtained by putting $%
n_{p}=0$ (or $p=0$) in above equation to obtain, 
\begin{equation}
\omega ^{2}=\frac{v_{Ai}^{2}k_{z}^{2}}{\left( 1+\lambda
_{e}^{2}k_{x}^{2}\right) }  \label{e11}
\end{equation}
which is the same linear dispersion relation of Inertial or slow shear
Alfv\'{e}n wave (SSAW) in e-i plasmas which has already been studied in Ref. 
\cite{r2}. It can be seen from Eq.(\ref{e10}) that the wave dispersion due
to electron inertial length in the presence of positrons is modified because
of the factor $p$. So the positron density can have significant effect on
the wave dynamics in nonlinear regime as well.

It may be mentioned here that Eq.(\ref{e10}) cannot reduce to the linear
dispersion relation of shear Alfv\'{e}n waves of e-p plasmas because the
polarization drifts of electrons and positrons have been ignored in the
limit $|\partial _{t}|<<\Omega _{j}$ (where $\Omega _{j}=\frac{q_{j}B_{0}}{%
m_{j}c}$ is the gyrofrequency of $j^{th}$ species and $j=e,p$).

\begin{center}
III. {\Large NONLINEAR SOLUTION}
\end{center}

We are interested in the stationary localized planar solution of the
non-linear set of equations. So we transform the set of Eqs. (1)-(8) in
moving frame $\xi $ defined as , 
\begin{equation}
\xi =K_{x}x+K_{z}z-\Omega t  \label{e12}
\end{equation}
where '$\Omega $' is the velocity of the non-linear structure in the moving
frame, $K_{x}$ and $K_{z}$ are the direction cosines in the x and z
directions, respectively and $K_{x}^{2}+$ $K_{z}^{2}=1$.

Now the electron continuity equation can be written as, 
\begin{equation}
v_{ez}=u\left( 1-n_{e}^{-1}\right)  \label{e13}
\end{equation}
the electron momentum equation can be written by using above equation, 
\begin{equation}
\frac{1}{K_{z}}\frac{\partial \varepsilon _{z}}{\partial \xi }=-u^{2}\frac{%
\partial ^{2}}{\partial \xi ^{2}}n_{e}^{-1}-\frac{1}{2}u^{2}\frac{\partial
^{2}}{\partial \xi ^{2}}\left( 1-n_{e}^{-1}\right) ^{2}  \label{e14}
\end{equation}
The normalized electron density is $n_{e}=\frac{n_{e}}{n_{e0}}$ and $%
\varepsilon _{z}=\frac{eE_{z}}{m}$ (where $m_{e}=m_{p}=m$). In order to
obtain Eq.(\ref{e13}), we have used the boundary conditions i.e., as $\xi
\rightarrow |\pm \infty |$ , $n_{e}\rightarrow 1$ and $v_{ez}\rightarrow 0$.

Similarly from positron continuity equation we have, 
\begin{equation}
v_{pz}=u\left( 1-n_{p}^{-1}\right)  \label{e15}
\end{equation}
and the positron momentum equation and the above relation give, 
\begin{equation}
\frac{1}{K_{z}}\frac{\partial \varepsilon _{z}}{\partial \xi }=u^{2}\frac{%
\partial ^{2}}{\partial \xi ^{2}}n_{p}^{-1}+\frac{1}{2}u^{2}\frac{\partial
^{2}}{\partial \xi ^{2}}\left( 1-n_{p}^{-1}\right) ^{2}  \label{e16}
\end{equation}
where $u=\frac{\Omega }{K_{z}}$ and normalized positron density is $n_{p}=%
\frac{n_{p}}{n_{p0}}$. We have again used the boundary conditions i.e., as $%
\xi \rightarrow |\pm \infty |$, $n_{p}\rightarrow 1$ and $v_{pz}\rightarrow
0 $ to obtain Eq.(\ref{e15}).

The ion continuity equation yields, 
\begin{equation}
-K_{x}\frac{\partial \varepsilon _{x}}{\partial \xi }=\Omega _{e}\Omega
_{i}\left( 1-n_{i}^{-1}\right)  \label{e17}
\end{equation}
where normalized ion density $n_{i}=\frac{n_{i}}{n_{i0}}$ , $\varepsilon
_{x}=\frac{eE_{x}}{m_{i}}$ and $\Omega _{e}=\frac{eB_{0}}{mc}$ (electron
gyrofrequency).

Using Eqs.(\ref{e13}) and (\ref{e15}) in Eq.(\ref{e8}), one can write
Ampere's law in $\xi $ co-ordinate as, 
\begin{equation}
K_{x}\frac{\partial B_{y}}{\partial \xi }=\frac{4\pi en_{e0}}{c}u\left[
p\left( 1-n_{p}^{-1}\right) n_{p}-n_{e}\left( 1-n_{e}^{-1}\right) \right]
\label{e18}
\end{equation}
Now after transforming Eq.(\ref{e7}) in $\xi $ co-ordinate and using the
above relation we have, 
\begin{equation}
K_{x}K_{z}^{2}\frac{\partial \varepsilon _{x}}{\partial \xi }-K_{x}^{2}K_{z}%
\frac{\partial \varepsilon _{z}}{\partial \xi }=\Omega ^{2}\lambda
_{e}^{-2}\ \left[ p\left( 1-n_{p}^{-1}\right) n_{p}-n_{e}\left(
1-n_{e}^{-1}\right) \right]  \label{e19}
\end{equation}
Equating left hand side (L.H.S) of Eqs.(\ref{e14}) and (\ref{e16}) and then
integrating the resulting equations twice w.r.t. $\xi $ we obtain, 
\begin{equation}
\left( v_{ez}-u\right) ^{2}+\left( v_{pz}-u\right) ^{2}=2u^{2}  \label{e20}
\end{equation}
Using Eqs. (\ref{e13}) and (\ref{e15}) in the above relations, we obtain, 
\begin{equation}
n_{p}^{-2}=2-n_{e}^{-2}  \label{e21}
\end{equation}
The quasi-neutrality yields, 
\begin{equation}
n_{i}=\frac{1}{\left( 1-p\right) }\left( n_{e}-pn_{p}\right)  \label{e22}
\end{equation}
Note that the above equations hold for $0\leqslant p<1$ in three component
e-p-i plasmas.

Eq. (\ref{e21}) and Eq. (\ref{e22}) yields, 
\begin{equation}
n_{i}^{-1}=\frac{\left( 1-p\right) }{\left[ n_{e}-\frac{p}{\sqrt{2-n_{p}^{-2}%
}}\right] }  \label{e23}
\end{equation}
Then Eq.(\ref{e21}) along with Eqs.(\ref{e14}) and (\ref{e16}) gives, 
\begin{equation*}
-K_{z}^{2}\left[ \Omega _{e}\Omega _{i}\left( 1-n_{i}^{-1}\right) \right]
+u^{2}K_{z}^{2}K_{x}^{2}\left[ \frac{\partial ^{2}}{\partial \xi ^{2}}%
n_{e}^{-1}+\frac{1}{2}\frac{\partial ^{2}}{\partial \xi ^{2}}\left(
1-n_{e}^{-1}\right) ^{2}\right]
\end{equation*}
\begin{equation}
=\Omega ^{2}\lambda _{e}^{-2}\ \left[ p\left( n_{p}-1\right) -\left(
n_{e}-1\right) \right]  \label{e24}
\end{equation}
Differentiating w.r.t $\xi $ twice the above equation and then after
simplification we obtain, 
\begin{equation}
K_{x}^{2}\lambda _{e}^{2}\frac{\partial ^{2}}{\partial \xi ^{2}}\left[
3n_{e}^{-4}\left( \frac{\partial n_{e}}{\partial \xi ^{\ }}\right)
^{2}-n_{e}^{-3}\frac{\partial ^{2}n_{e}}{\partial \xi ^{2\ }}\right] -\frac{%
\left( 1-p\right) }{M^{2}}\frac{\partial ^{2}}{\partial \xi ^{2}}\left(
1-n_{i}^{-1}\right) =\frac{\partial ^{2}}{\partial \xi ^{2}}\left(
pn_{p}-n_{e}\right)  \label{e25}
\end{equation}
where $M=\frac{u}{v_{A}}$ is defined as Mach number.

Integrating Eq.(\ref{e25}) twice w.r.t. $\xi $, we have, 
\begin{equation}
K_{x}^{2}\lambda _{e}^{2}\left[ 3n_{e}^{-4}\left( \frac{\partial n_{e}}{%
\partial \xi ^{\ }}\right) ^{2}-n_{e}^{-3}\frac{\partial ^{2}n_{e}}{\partial
\xi ^{2\ }}\right] -\frac{\left( 1-p\right) }{M^{2}}\left(
1-n_{i}^{-1}\right) =\left( n_{p}p-n_{e}\right) +\left( 1-p\right)
\label{e26}
\end{equation}
In order to obtain above equation, we have used the boundary conditions
i.e., $n_{p}\rightarrow 1$ $n_{i}\rightarrow 1$ and $n_{e}\rightarrow 1$as $%
\xi \rightarrow |\pm \infty |$.

Let us define $R=n_{e}^{-3}\frac{\partial n_{e}}{\partial \xi }$ , then
multiplying Eq.(\ref{e26}) by '$R$' both sides and after integrating once
w.r.t $\xi $ we obtain, 
\begin{equation}
\frac{1}{2}\left( \frac{\partial n_{e}}{\partial \xi }\right) ^{2}+V\left(
n_{e}\right) =0  \label{e27}
\end{equation}
The Sagdeev potential is defined as, 
\begin{equation*}
V\left( n_{e}\right) =\frac{n_{e}^{6}}{K_{x}^{2}}\left[ \frac{1}{n_{e}}(1+p%
\sqrt{2n_{e}^{2}-1})-\frac{(1-p)}{2n_{e}^{2}}\left( \frac{1}{M^{2}}+1\right)
-\frac{(1-p)^{2}}{M^{2}}\int_{1}^{n_{e}}\frac{1}{n_{e}^{4}\left[ 1-\frac{p}{%
\sqrt{2n_{e}^{2}-1}}\right] }dn_{e}\right.
\end{equation*}
\begin{equation}
\left. -\frac{1}{2}(1+3p)+\frac{1}{2}\frac{(1-p)}{M^{2}}\right]  \label{e28}
\end{equation}
where $\xi =\frac{\xi }{\lambda _{e}}$ has been normalized. We have used the
boundary conditions i.e., as $\xi \rightarrow |\pm \infty |$ then $\left( 
\frac{\partial n_{e}}{\partial \xi }\right) \rightarrow 0$ and $%
n_{e}\rightarrow 1$ to obtain Eq.(\ref{e27}).

Equation(\ref{e27}) is a well known equation in the form of ''energy
integral'' of an oscillating particle of a unit mass, with velocity $(\frac{%
\partial n_{e}}{\partial \xi })$ and position $n_{e}$ in a potential $%
V\left( n_{e}\right) $. The conditions for the existence of localized
solution of Eq.(\ref{e27}) require that i) $V(1)=V(N_{0})=\frac{\partial V}{%
\partial n_{e}}\mid _{n_{e}=1}=0$, ($N_{0}$ is the point where the curve
crosses the $n_{e}$ axis and it can have values $>$ or $<1$) and it
represents the maximum amplitude of the soliton ii)$\frac{\partial ^{2}V}{%
\partial n_{e}^{2}}\mid _{n_{e}=1}<0$ (where $n_{e}=1$ is the unstable
point) and from second condition it is seen that solitary structures are
formed only in the super Alfv\'{e}nic region.

The limiting case of two component e-i plasma can be obtained by putting $%
p=0 $ in Eq.(\ref{e27}) and the Sagdeev potential in this case turn out to
be, 
\begin{equation}
V\left( n\right) =\frac{n^{6}}{b}\left[ \frac{1}{2}\left( \frac{1}{M^{2}}%
+1\right) \left( 1-\frac{1}{n^{2}}\right) +\frac{1}{n}-1+\frac{1}{3M^{2}}%
\left( \frac{1}{n^{3}}-1\right) \right]  \label{e29}
\end{equation}
which is the same as obtained by Shukla et. al., \cite{r2} and $b=\lambda
_{e}^{2}K_{x}^{2}$ has been defined.

\begin{center}
IV. {\Large NUMERICAL SOLUTIONS}
\end{center}

The numerical solutions of Eq.(\ref{e27}) are obtained for solitary
structures in the absence as well as in the presence of positrons in e-i
plasmas. The Sagdeev potential in the presence of positron (as a third
species) in two component e-i plasmas becomes complicated and the third
integral term does not have simple analytical solution. However, the
numerical solutions exist and the plots of the Sagdeev potential '$V$' vs
normalized electron densities '$n_{e}$' corresponding to $p=0.2$ (dashed
curve) i.e., for e-p-i plasmas as well as corresponding to $p=0$ (solid
curve) i.e., e-i plasmas for $M=1.2$, $K_{x}=0.1$ have been shown in Fig.1.
The corresponding electron density dips have been shown in Fig.2 for the
same parameters as given in Fig.1. It can be seen from the figures that the
amplitude and the width of solitary structures reduces with increase in
percentage of positrons in e-i plasmas. The normalized density profiles of
three species electron-positron-ion plasmas are shown in Fig. 3.

The effects on the direction of propagation on the solitons for both the
cases e-i and e-p-i plasmas are shown in figures 4 and 5, respectively. The
plots for different directions of propagation i.e., for $K_{x}=0.1$ (solid
curve) and $K_{x}=0.3$ (dotted curve) for two component e-i ( $p=0$) plasmas
and for $K_{x}=0.1$ (dotted curve) and $K_{x}=0.3$ (solid curve) in three
component e-p-i ($p=0.2$) plasmas for the same value of $M=1.2$ have been
shown in figures 3 and 4. It can be seen from the figures that propagation
direction effects on the width of the solitary structures. The width of the
solitary structure increases with the increase in the obliqueness of the
wave.

The dependence of Mach number on soliton corresponding to $M=1.6$ (solid
curve) and $M=1.2$ (dotted curve) for same $p=0.2$, $K_{x}=0.1$ has been
shown in Fig.6. It can be seen from the figure that the Mach number has
significant effect on the amplitude as well as on the width of the solitary
structure. The wave amplitude increases and the width decreases with the
increase in Mach number.

\begin{center}
V. {\Large DISCUSSION ON RESULTS}
\end{center}

We have studied the solitary pulse formation of inertial or slow shear Alfv%
\'{e}n waves in electron-positron-ion plasmas. The conditions on the Mach
number for the formation of such nonlinear structures in e-i plasmas were
presented long ago \cite{r2}. It was predicted that super Alfv\'{e}nic
density depletion regions can be formed in e-i plasmas corresponding to an
arbitrary amplitude perturbation. However the density profiles were not
plotted and the dependence of the nonlinear structures on the propagation
direction was not investigated. We have noticed that the pulse width
increases with the increase in obliqueness (with respect to the external
magnetic field) of the propagation direction in e-i plasmas as shown in
Fig.4.

It is pointed out that the conditions for the formation of solitary pulses
in e-p-i plasmas due to large amplitude SSAW perturbation are similar to the
e-i case. That is the density depletion regions can be formed in the super
Alfv\'{e}nic region i.e, for $1<M$. But the amplitude of the soliton
decreases with the increase in the number density of positrons. This result
is very common in e-p-i plasmas. For example the amplitudes of the solitary
ion acoustic \cite{r5} and solitary kinetic Alfv\'{e}n waves\cite{r10} also
decrease with the increase in the concentration of positrons in e-i plasmas.

In principle, the Alfv\'{e}n waves can have a wide range of temporal and
spatial scales in e-p-i plasmas because the frequency of the linear wave can
vary from $\omega =v_{Ai}k_{z}$ (with $\omega <\Omega _{i}$) to $\omega
=v_{Ape}k_{z}$ (with $\omega <\Omega _{e,p}$), where $v_{Aep}=(\frac{%
B_{0}^{2}}{8\barwedge n_{0}m})^{\frac{1}{2}}$ (where $m_{e}=m_{p}=m$)\ is
the Alfv\'{e}n wave speed, and correspondingly the spectrum of wavelengths
can be broader. These variations in the temporal and spatial scales depend
upon the concentration ratios of different species. Therefore it seems
interesting to analyze the linear and nonlinear wave propagation in such
plasmas.

Here, we have studied only the Inertial or slow shear Alfv\'{e}n waves in
e-p-i plasmas which propagate on ionic time scale. It may be mentioned that
the limiting case of e-p plasmas can not be obtained from our equations
because we have ignored the polarization drifts of electrons and positrons.
Our findings can be useful to explain some aspects of laboratory and
astrophysical space e-p-i plasmas situations.

\textbf{Acknowledgments:}

One of the author (SM) is grateful to the Organizers of the ''12th
International Congress on Plasma Physics'' for providing half air travel and
financial support for local hospitality to participate in the Conference and
also thankful to ''Organization of Islamic Conference Standing Committee on
Scientific and Technological Cooperation (COMSTECH), Islamabad Pakistan''
for supporting rest of the half air travel.

\begin{itemize}
\item  \textbf{Fig.1:}The Sagdeev potential '$V$' is plotted against
electron density '$n_{e}$' for $M=1.2$, $K_{x}=0.1$, $p=0$ (solid curve) and 
$p=0.2$ (dotted curve).

\textbf{Fig.2:}The normalized electron density dip decreases with the
increase in positrons in e-p-i plasmas for the same parameters a given in
Fig.1.

\textbf{Fig.3:} The normalized density profiles of electron, positron and
ion are plotted for $p=0.2$ , while $M$ and $K_{x}$ corresponds to the same
values as given in Fig.1.

\textbf{Fig.4:}Effect of propagation direction on the width of the solitary
pulse in e-p-i plasmas is shown with $K_{x}=0.1$ (solid curve) and $%
K_{x}=0.3 $ (dotted curve) for $M=1.2$ and $p=0$.

\textbf{Fig.5:}Effect of propagation direction on the width of the solitary
pulse in e-p-i plasmas is shown with $K_{x}=0.1$ (dotted curve) and $%
K_{x}=0.3$ (solid curve) for $M=1.2$ and $p=0.2$.

\textbf{Fig.6:}The effect of Mach number on the solitons in e-p-i plasmas is
shown plotting $n_{e}$ and $\xi $ for $p=0.2$, $K_{x}=0.1$, $M=1.6$ (solid
curve) and $M=1.2$ (dotted curve)
\end{itemize}

\end{document}